\documentclass[epj,twocolumn]{webofc}
\usepackage[varg]{txfonts}   % Web of Conferences font
\woctitle{19th International Seminar on High Energy Physics
"QUARKS-2016"}
\begin{document}
\title{Relativistic constituent model in sector of light mesons}
%
% subtitle is optionnal
%
%%%\subtitle{Do you have a subtitle?\\ If so, write it here}

\author{\firstname{A.F.} \lastname{Krutov}\inst{1}\fnsep\thanks{\email{krutov@samsu.ru}} \and
        \firstname{R.G.} \lastname{Polezhaev}\inst{1}\fnsep\thanks{\email{polezaev@list.ru}}
        \and
        \firstname{V.E.} \lastname{Troitsky}\inst{2}\fnsep\thanks{\email{troitsky@theory.sinp.msu.ru}}
        % etc.
}

\institute{Samara University, 443086 Samara, Russia \and
           D.V.~Skobeltsyn Institute of Nuclear Physics,
           Moscow State University, Moscow 119991, Russia
%           \and
%           Last address
          }

\abstract{%
We present a brief survey of some results on electroweak
properties of composite systems that are obtained in the
frameworks of our version of the instant form of relativistic
quantum mechanics (RQM). Our approach describes well the  $\pi$-
and the $\rho$- mesons in wide range of momentum transfers
$Q^{2}$. At large $Q^{2}$ the obtained pion form factor
asymptotics coincides with that of QCD predictions. The method
permits to perform analytic continuation of pion form factor to
complex plane of momentum transfers that is in accordance with
predictions of quantum field theory.}
\maketitle
\section{Introduction}
\label{intro}

The purpose of this paper is to present a version of relativistic
composite quark model developed by the authors (see,
e.g.,\cite{KrT09}) for the study of composite systems of light
quarks.The investigation of electroweak properties of light mesons
is an important part of  the study of the transition region where
the perturbative QCD behavior starts to be valid. In this
connection such particles are in the focus of experiments on
up-to-day  accelerators. The most important results here are the
following: measurement of the pion form factor at large momentum
transfers in JLab (see, e.g., \cite{Blo08,HuB08}), the obtaining
of the $\rho$-meson decay constant from the reaction
$\tau\to\rho\,\nu_\tau$ \cite{MeS15,Oli14}, the magnetic moment
measurement of the radiation transition $\rho\to\pi\gamma^*$. The
interesting result obtained by BABAR collaboration concerning the
deviation of the behavior of the transition form factor
$F_{\pi\rightarrow\gamma\gamma^{*}}(Q^2)$ for large momentum
transfer from that predicted by perturbative QCD \cite{Aub09}
remains unexplained.

% PEPAN 2009
The theoretical description of systems containing light quarks,
requires the account of relativistic effects even at low energies.
It should be underlined however, that quantitative description of
relativistic composite hadronic systems is a very complicated
problem which can hardly be solved completely in the near future,
since for this purpose it is necessary to solve the many-body
relativistic problem, sometimes, with an interaction which is not
always known well. The application of methods of the field theory
for solution of this problem encounters serious difficulties.
Thus, for example, it is known that perturbative quantum
chromodynamics cannot be applied to the problem of the bound
states of quarks (see, e.g.,~\cite{Gro93,Kei94W}). In this regard,
the so called relativistic constituent models became widely used
for the description of composite hadronic systems.

One of the main problem in construction of these models is known
to be the problem of construction of operators of transition
currents. Generally speaking, the complexity of the construction
of, for example, the operator of electromagnetic current of the
composite system satisfying the Lorentz-covariance and
conservation conditions appears in all approaches, including the
perturbative quantum field theory. Thus, to ensure the
conservation law in the framework of the Bethe-Salpeter equation
and quasipotential equations it is necessary to go beyond the
framework of the impulse approximation (IA), i.e., it is necessary
to add the so called two-particle currents (see, e.g.,
\cite{CoR94}) to the current operator; these currents are
interpreted, for example, in nuclear physics as exchange meson
currents. It should be noted that the approach to describing the
electroweak structure of two-particle composite systems presented
in this survey has the following characteristic features:  the
matrix element of the electroweak current of the composite system
automatically satisfies the relativistic covariance conditions;
the matrix element of the electromagnetic current satisfies the
conservation law; the IA is formulated in a relativistically
invariant way and in the case of the electromagnetic current and
with account of the conservation law, the so called modified
impulse approximation (MIA) is formulated. This procedure of
construction of the current operators actually realizes the
Wigner-Eckart theorem on the Poincar\'e group, i.e., it allows
separating from the matrix element of the operator any tensor
dimension of the reduced matrix elements (form factors) which are
invariants under the Poincar\'e group transformations. In general,
these form factors are not classical but generalized functions.

%PEPAN 2009
The method in the relativistic theory of composite systems which
will be used here is based on the direct realization of the
Poincar\'e algebra on the set of dynamic observable systems and
dates back to the P.Dirac's paper \cite{Dir49}. This approach is
called the theory of direct interaction or the relativistic
quantum mechanics with fixed number of particles (RQM) (see,
e.g.,review \cite{KrT09} and references therein). From the point
of view of the principles underlying it, the RQM occupies the
intermediate position between the local quantum field theory and
nonrelativistic quantum mechanical models.

It should be noted that the field theory and the RQM are
formulated as fundamentally different structures and the
establishment of a connection between them is a complicated
problem that has not been solved yet. Unlike the field theory, in
the Poincar\'e-invariant quantum mechanics, the finite number of
degrees of freedom are separated initially, that represents some
model. The covariance of description in the RQM is provided by the
construction on the Hilbert state space of the composite system
with the finite number of degrees of freedom of the unique unitary
transformation of the inhomogeneous group $SL(2,C)$, which is the
universal covering of the Poincar\'e group. In this case, the
interaction is included in the group generators (operators of
observable systems).

The RQM can be realized using different methods and in different
forms of dynamics (instant form, point form, light-front form)
which differ by methods of inclusion of interaction to the algebra
of group generators.

In this survey basic attention was paid to the description of the
electroweak structure of composite systems in the framework of the
instant form of RQM.

\section{Instant form of RQM}
\label{sec-1}

Relativistic invariance means that on the Hilbert
state space of the system the unitary representation of the
Poincar\'e group (or, more precisely, of the inhomogeneous group
$SL(2,C)$ which is the universal covering of the Poincar\'e group)
~\cite{Nov72} is realized.

Considering infinitesimal transformations and introducing the
generators of translations $\hat P\,^\mu$ and space-time rotations
$\hat M\,^{\mu\nu}$, we arrive at the Poincar\'e algebra in the
common way,
$$
[\hat M^{\mu \nu }, \hat P^\sigma ] = -i(g^{\mu \sigma }\hat P^\nu
- g^{\nu  \sigma }\hat P^\mu )\;,\quad
$$
$$
[\hat M^{\mu \nu },\hat M^{\sigma \rho }] = -i(g^{\mu \sigma }\hat
M^{\nu \rho } - g^{\nu \sigma }\hat M^{\mu \rho }) - (\sigma
\leftrightarrow \rho ),
$$
\begin{equation}
[\hat P^\mu, \hat P^\nu ] = 0. \label{algebra}
\end{equation}
In (\ref{algebra}) $g^{\mu\nu}$ is the metric tensor in the
Minkowsky space.

The construction of the representation of the Poincar\'e group in
the Hilbert space is reduced to finding the generators $\hat
P^\mu$, $\hat M^{\mu \nu}$ in terms of dynamic variables of the
system. In the case of the system of noninteracting particles
generators in (\ref{algebra}) have the clear physical meaning,
$\hat P^0 \equiv \hat H$ is the operator of total energy,$\hat
{\vec P} = (\hat P^1,\hat P^2,\hat P^3)$ is the operator of total
3- momentum, $\hat {\vec J} = (\hat M^{23},\hat M^{31},\hat
M^{12})$ is the operator of total angular momentum, and $\hat
{\vec N} = (\hat M^{01},\hat M^{02},\hat M^{03})$ are the Lorentz
boost generators. However, inclusion of the interaction between
particles in this approach involves some problems; the essence of
these problems can be illustrated by considering first the quantum
nonrelativistic theory and its invariance group, the Galilean
group. After the known transition to the central extension of the
Galilean group to the covering group $SU(2)$ \cite{FuN90} we
obtain the 11-parametric group with the set of generators
$$
\hat H\;,\;\hat{\vec P}\;,\;\hat{\vec J}\;,\;\hat{\vec K}\;,\;\hat
M\;,
$$
where $\hat{\vec K}$ are the generators of the Galilean boosts and
$\hat M$ is the mass operator. The other generators coincide with
the corresponding operators of the Poincar\'e group.

The following nonzero commutation relations are contained in the
Galilean algebra:
$$
[\hat J_i,\hat J_j] = i\epsilon_{ijk}\,\hat J_k\;,\quad [\hat
J_i,\hat K_j] = i\epsilon_{ijk}\,\hat K_k\;,\quad
$$
$$
[\hat J_i,\hat P_j] = i\epsilon_{ijk}\,\hat P_k\;,
$$
\begin{equation}
[\hat K_i,\hat H] = -i\,\hat P_i\;,\quad [\hat K_i,\hat P_k] =
-i\delta_{ik}\,\hat M\;. \label{algG}
\end{equation}

In nonrelativistic quantum mechanics the operator of interaction
is added to the operator of total energy, $ \hat H\;\;\to \hat H +
\hat V\;.$ In order to preserve the Galilean invariance of the
theory, i.e., to preserve algebra (\ref{algG}) under such
re-definition of the operator of total energy, the following
conditions should be imposed on the operator of interaction:
\begin{equation}
[\hat{\vec P},\hat V] = [\hat{\vec J},\hat V] =
[\vec\bigtriangledown_P,\hat V] = [\hat M,\hat V] = 0\;.
\label{consG}
\end{equation}
Since the generator $\hat H$ is absent in the right-hand sides of
relations (\ref{algG}), it is not necessary to include the
interaction to other generators of the group in order to preserve
the Galilean algebra.

The matter is different in the case of the Poincar\'e group. Let
us consider one of the generators of algebra (\ref{algebra}) (see,
e.g., \cite{Kei94W}):
\begin{equation}
[\hat P^j \hat N^k ] = i\,\delta^{jk}\,\hat H\;. \label{key}
\end{equation}
At inclusion of the interaction to the operator of total energy
described above, the right-hand side of (\ref{key}) depends on the
interaction; therefore, either both generators in the left-hand
side of (\ref{key}), or one of them should depend on the
interaction. Thus, in order to preserve commutation relations in
(\ref{algebra}) it is necessary to make other generators in set
(\ref{algebra}) dependent on the interaction. The generators of
the algebra are separated into two types in this case: generators
independent of interaction which form the so called kinematical
subgroup and generators depending on interaction, Hamiltonians.
The separation of the generators into kinematical generators and
Hamiltonians is not unambiguous. Different methods for the
separation of the kinematical subgroup result in different forms
of dynamics. Usually three basic forms of dynamics are identified:
the point form, the instant form, and the light-front form.

Further, the instant form of dynamics will be used in which the
kinematical subgroup is comprised of the generators of the group
of rotations and shifts of the Euclidean space, $ \hat{\vec
J}\;,\quad\hat{\vec P}\;$, the other generators are the
Hamiltonians, i.e., depend on the interaction, $ \hat P^0\;,\quad
\hat{\vec N}\;.$

One of the technical methods for inclusion of interaction to
algebra (\ref{algebra}) allowing to preserve commutation relations
is the additive inclusion of interaction to the mass operator, the
so called Bakamajian-Thomas  procedure \cite{BaT53} (see also
\cite{KrT09}):
\begin{equation}
\hat M_0 \to \hat M_I = \hat M_0 + \hat V \;. \label{M0toMI}
\end{equation}
Here, $\hat M_0$ is the operator of the invariant mass of the
system without interaction, $\hat M_I$ is the mass operator of the
system with interaction. In the instant form of dynamics the
operator of interaction should satisfy the following conditions:
\begin{equation}
\hat M_I = \hat M_I^+\;,\quad \hat M_I\;>\;0\;, \label{MI+}
\end{equation}
\begin{equation}
\left [\hat {\vec P},\,\hat V\right ] = \left [\hat {\vec
J},\,\hat V\right ] = \left [ \vec\bigtriangledown_P,\,\hat
V\right ] = 0\;. \label{[PU]=0}
\end{equation}
Conditions (\ref{MI+}) represent the spectral conditions for the
mass operator. Equalities (\ref{[PU]=0}) provide the satisfaction
of algebraic relations (\ref{algebra}) in the system with an
interaction. Relations (\ref{[PU]=0}) are not too limiting, for
example, any nonrelativistic interaction potential of particles
(\ref{consG}) satisfies these relations.

After the operator $\hat V$ is defined, the interaction can be
introduced in a different way,
\begin{equation}
\hat U = (1/4)({\hat M}_I^2 - {\hat M}_0^2) = (1/4)({\hat V}^2 +
[\hat M_0,\,\hat V]_+)\;. \label{VI}
\end{equation}
Interaction (\ref{VI}) is introduced from the considerations of
convenience, since in this case the problem of finding the
eigenvalue of the mass operator can be represented in the form
similar to the nonrelativistic Schr\"odinger equation (see, e.g.,
\cite{KeP91}). Operator (\ref{VI}) also satisfies conditions
(\ref{[PU]=0}) by definition.

The wave function of the system of interacting particles in the
RQM is determined as the eigenfunctions of the total commuting set
of operators. In the instant form of dynamics this set consists of
the operators
\begin{equation}
 {\hat M}_I^2\;(\hbox{or}\;\hat M_I)\;,\quad
{\hat J}^2\;,\quad \hat J_3\;,\quad \hat {\vec P}\;.
\label{complete}
\end{equation}
${\hat J}^2$ is the operator of the total angular momentum
squared. In the instant form of dynamics the operators ${\hat
J}^2\;,\; \hat J_3\;,\; \hat {\vec P}$ coincide with the
corresponding operators of the system without interaction. Thus,
only the operator $\hat M_I^2\;(\hat M_I)$ depends on the
interaction in (\ref{complete}).

It is possible to introduce a basis in which the motion of the
center of mass of two particles is separated and three of four
operators of set (\ref{complete}) are diagonal:
$$
|\,\vec P,\;\sqrt {s},\;J,\;l,\;S,\;m_J\,\rangle\;,
$$
$$
\langle\,\vec P,\;\sqrt {s},\;J,\;l,\;S,\;m_J |\,\vec P\,',\;\sqrt
{s'},\;J',\;l',\;S',\;m_{J'}\,\rangle =
$$
\begin{equation}
= N_{CG}\,\delta^{(3)}(\vec P - \vec P\,')\delta( \sqrt{s} -
\sqrt{s'})\delta_{JJ'}\delta_{ll'}\delta_{SS'}\delta_{m_Jm_{J'}}\;,
\label{PkJlSm}
\end{equation}
$$
N_{CG} = \frac{(2P_0)^2}{8\,k\,\sqrt{s}}\;,\quad k =
\frac{\sqrt{\lambda(s\,,\,M^2\,,\,M^2)}}{2\,\sqrt{s}}\;,
$$
where $P_\mu = (p_1 +p_2)_\mu$, $P^2_\mu = s$, $\sqrt {s}$ is the
invariant mass of the system of two particles, $l$ is the orbital
momentum in the center-of-mass system, $\vec S\,^2=(\vec S_1 +
\vec S_2)^2 = S(S+1)\;,\;S$ is the total spin in the
center-of-mass system, $J$ is the total angular momentum, $m_J$ is
the projection of the total angular momentum, $M$ is the mass of
the constituents, and  $\lambda(a,b,c) = a^2 + b^2 + c^2 - 2(ab +
ac + bc)$.

Basis (\ref{PkJlSm}) diagonalizes the operators ${\hat J}^2\;,\;
\hat J_3\;,\; \hat {\vec P}$ in (\ref{complete}). Thus, the
problem of calculation of the wave function of the system is
reduced to the diagonalization of the operator $\hat M_I^2$ (or
$\hat M_I$).

It should be noted that the eigenvalue problem for the operator
$\hat M_I^2$ has the form of the nonrelativistic Schr\"odinger
equation (see, e.g., review \cite{KrT09}). Thus, the operator can
be considered as the phenomenological nonrelativistic potential.
%\subsection{Subsection title}

\section{Construction of the electroweak current matrix elements}
\label{sec-2}

%TMPH Wigner-Eckard
Let us describe now a general
method of canonical parameterization of local operator matrix
elements briefly (see  e.g., \cite{KrT09, KrT05} for details).

The main idea of the parametrization can be formulated as follows.
Using the variables entering the state vectors which define the
matrix elements one has to construct two types of objects.

1. A set of linearly independent matrices which are Lorentz
scalars (scalars or pseudoscalars). This set describes transition
matrix elements non-diagonal in spin projections in the initial
and finite states, as well as the properties defined by the
discrete space--time transformations.

2. A set of linearly independent objects with the same tensor
dimension as the operator under consideration (for example,
four--vector, or four--tensor of some rank). This set describes
the matrix element behavior under the action of Lorentz group
transformations.

The operator matrix element is written as a sum of all possible
objects of the first type multiplied by all possible objects of
the second type. The coefficients in this representation as a sum
are just the reduced matrix elements -- form factors.

The obtained representation is then modified with the use of
additional conditions for the operator, such as the conservation
laws, for example. In order to satisfy these additional conditions
in some cases some of the coefficients (form factors) occur to be
zero.

To demonstrate this let us consider the parameterization of the
matrix elements taken between the states of a free particle of
mass $M$ in different simple cases. Let us normalize the state
vectors as follows:
\begin{equation}
\langle\,\!\vec p\,,m\,|\,\vec p\,'\,,m'\,\!\rangle =2p_0\,\delta
(\vec p - \vec p\,')\,\delta _{mm'}\;, \label{normg}
\end{equation}
here $\vec p,\vec p\,'$ are three--momenta of particle,$p_0 =
\sqrt{M^2 + \vec p\,^2}$, $m,m'$ are spin projections.

Let us consider now the 4-vector operator $j_\mu(0)$. To
parametrize the matrix element one needs a set of quantities of
the appropriate tensor dimension. Using the variables entering the
particle state vectors one can construct one pseudovector of the
covariant spin operator $\Gamma^\mu (p')$ (see, e.g.,
\cite{KrT09}):
$$
\Gamma_0(p) = (\vec p\vec j)\;,\quad \vec \Gamma(p) =  M\,\vec j +
\frac {\vec p(\vec p\vec j)}{p_0 + M}\;, \quad
$$
\begin{equation}
\Gamma^2 = -M^2\,j(j+1)\; \label{Gamma mu}
\end{equation}
and three independent vectors:
$$
K_\mu = (p - p')_\mu = q_\mu\,,\quad K'_\mu = (p + p')_\mu
\,,\quad
$$
\begin{equation}
R_\mu =\epsilon _{\mu \,\nu \,\lambda\,\rho}\, p^\nu
\,p'\,^\lambda\, \Gamma^\rho (p')\;. \label{kk'RG}
\end{equation}
Here $\epsilon _{\mu \,\nu \,\lambda\,\rho}$ is a completely
anti-symmetric pseudo-tensor in four dimensional space-time with
$\epsilon _{0\,1\,2\,3}= -1$.

The set in question of linearly independent matrices in spin
projections of the initial and the final states giving the set of
independent Lorentz scalars is presented by $2j + 1$ quantities:
\begin{equation}
D^j(p,\,p')\,(p_\mu \Gamma^\mu (p'))^n\;,\quad n = 0,1,\ldots
,2j\;. \label{pseud}
\end{equation}

The operator matrix element contains the matrix elements of the
listed quantities multiplied by Wigner's rotation matrix
$D^j(p,\,p')$ from the left. Each of such products is to be
multiplied by the sum of linearly independent scalars
(\ref{pseud}):
$$
\langle\,\vec p,\,M,\,j,\,m\,|j_\mu(0)|\,\vec
p\,',\,M,\,j,\,m'\,\rangle =
$$
$$
= \sum_{m''}\,\langle\,m|D^j(p,\,p')|m''\rangle
\langle\,m''|\,F_1\,K'_\mu + F_2\,\Gamma_\mu (p') +
$$
\begin{equation}
F_3\,R_\mu + F_4\,K_\mu |m'\rangle\;, \label{<|j|>=F_is}
\end{equation}
where
\begin{equation}F_i = \sum _{n=0}^{2j}\,f_{in}(Q^2)(ip_\mu\Gamma^\mu(p'))^n\;.
\label{Fi}
\end{equation}

Let us impose some additional conditions on the operator:
self-adjointness, orthogonality of the vectors in the
parametrization (\ref{<|j|>=F_is}), parity conservation and the
conservation condition: $q_\mu\,j_\mu(0)=0$.

So, finally we have the parametrization of the matrix element
which has following form for particle with spin 1/2:
$$
\langle\,\vec p,\,M,\,\frac{1}{2},\,m\,|\,j_\mu(0)\, |\,\vec
p\,',\,M,\,\frac{1}{2},\,m'\,\rangle =
$$
$$
= \sum_{m''}\,\langle\,m|D^{1/2}(p,\,p')|m''\rangle
\langle\,m''|\,f_{10}(Q^2)\,K'_\mu +
$$
\begin{equation}
if_{30}(Q^2)\,R_\mu|m'\rangle\;, \label{<|j|>=K+R}
\end{equation}
The form factors $f_{10}(Q^2)$ and $f_{30}(Q^2)$ are the electric
and the magnetic form factors of the particle, respectively. These
form factors are connected with Sachs form factors $G_{E}(Q^2)$
and $G_{M}(Q^2)$:
$$
f_{10}(Q^2)=\frac{2M}{\sqrt{4M^2+Q^2}}G_{E}(Q^2)\;,\quad
$$
\begin{equation}\label{sak}
f_{30}(Q^2)=-\frac{4}{M\sqrt{4M^2+Q^2}}G_{M}(Q^2)\;.
\end{equation}

The developed procedure can be applied to the construction of the
electromagnetic current two-quark system. The following integral
representation for the pion form factor in the MIA (see,
e.g.\cite{KrT01,KrT02, KrT09}) holds:
\begin{equation}
F_\pi(Q^2)=\int
\mathrm{d}\sqrt{s}\,\mathrm{d}\sqrt{s'}\,\varphi(k)\,g_0(s,Q^2,s')\,\varphi(k')\;.
\label{ffpi}
\end{equation}
Here $\varphi(k)$ is pion wave function in the sense of RQM,
$g_0(s,Q^2,s')$ is the free two-particle form factor describing
the electromagnetic properties of the two noninteracting quarks
without interaction with the quantum numbers of the pion. It may
be obtained explicitly by the methods of relativistic kinematics
and is a relativistic invariant function.

The wave function in (\ref{ffpi}) has the following structure:
\begin{equation}
\varphi(k) = \sqrt[4]{s}\,u(k)k\;,\quad s=4(k^2+M^2)\;.
\label{phi}
\end{equation}
Below for the function $u(k)$ we use some phenomenological wave
functions.

The function  $g_0(s,Q^2,s')$ is written in terms of the quark
electromagnetic form factors in the form
$$
g_0(s,Q^2,s')=
  \frac{(s+s'+Q^2)Q^2}{2\sqrt{(s-4M^2)
(s'-4M^2)}}\;
$$
$$
\times \frac{\theta(s,Q^2,s')}{{[\lambda(s,-Q^2,s')]}^{3/2}}
\frac{1}{\sqrt{1+Q^2/4M^2}}
$$
$$
\times\left\{(s+s'+Q^2)[G^q_E(Q^2)+G^{\bar q} _E(Q^2)]\right.
$$
$$
\times\cos{(\omega_1+\omega_2)} + \frac{1}{M}\,\xi(s,Q^2,s')
(G^q_M(Q^2)
$$
\begin{equation}
\left.+ G^{\bar q}_M(Q^2))\sin(\omega_1+\omega_2)\right\}\;.
\label{g_0}
\end{equation}
Here $\xi=\sqrt{ss'Q^2-M^2\lambda(s,-Q^2,s')}$,

$\omega_1$ and $\omega_2$ are the Wigner rotation parameters:
$$
\omega_1\!=\!\arctan{\frac{\xi(s,Q^2,s')}{M[(\sqrt s\! +\! \sqrt
{s'})^2\! +\! Q^2]\! +\! \sqrt{ss'}(\sqrt s\! +\! \sqrt{s'})}},
$$
$$
\omega_2\! =\! \arctan{\frac{\alpha(s,s')\xi(s,Q^2,s')}{M(s\! +\!
s'\! +\! Q^2)\alpha( s,s')\! +\! \sqrt{ss'}( 4M^2\! +\! Q^2)}},
$$
$\alpha(s,s') =  2M+\sqrt s+\sqrt {s'}$, $\theta(s,Q^2,s')=
\vartheta(s'-s_1) -\vartheta(s'-s_2)$, $\vartheta$ is the step
function,
$$
s_{1,2}=2M^2+\frac{1}{2M^2} (2M^2+Q^2)(s-2M^2)
$$
$$
\mp \frac{1}{2M^2} \sqrt{Q^2(Q^2+4M^2)s(s-4M^2)}.
$$

Expressions similar to (\ref{ffpi}) take a place for charge $G_C(Q^2)$,
quadrupole $G_Q(Q^2)$ and magnetic $G_M(Q^2)$ form factors of the
$\rho$--meson:
$$
G_C(Q^2) = \int\,d\sqrt{s}\,d\sqrt{s'}\,
\varphi(s)\,g_{0C}(s\,,Q^2\,,s')\, \varphi(s')\;,
$$
\begin{equation}
G_Q(Q^2) = \frac{2\,M_c^2}{Q^2}\,\int\,d\sqrt{s}\,d\sqrt{s'}\,
\varphi(s)\,g_{0Q}(s\,,Q^2\,,s')\,\varphi(s')\;, \label{GqGRIP}
\end{equation}
$$
G_M(Q^2) =-\,M_c\,\int\,d\sqrt{s}\,d\sqrt{s'}\,
\varphi(s)\,g_{0M}(s\,,Q^2\,,s')\, \varphi(s')\;,
$$
where $\varphi(k)$ is the two-quarks wave function of the
$\rho$-meson in the sense of RQM, $g_{0C}\;,\;g_{0Q}\;,\;g_{0M}$
are the free two-particle form factors describing the
electromagnetic properties of the two noninteracting quarks
without interaction with the quantum numbers of the $\rho$-meson.
The explicit form of free two-particle form factors is cumbersome;
it can be found in Ref.~\cite{KrT02arx} which is an extended
version of Ref.~\cite{KrT03}.

The method of the parameterization of the current matrix elements
has been generalized in the paper \cite{KrP16} to the case of the
nondiagonal in the total angular momentum matrix elements. In
particular the expression for the lepton decay constant of the
$\rho$-meson was obtained in the 4-fermion approximation:
\begin{eqnarray}\label{decayconst111}
&&\hspace{-2mm}f_{\rho}=\frac{\sqrt{3}}{\sqrt{2}
\pi}\int^{\infty}_{0}\,dk\,k^2\,u(k) \frac{(\sqrt{k^2 + M^2} +
M)}{(k^2 + M^2)^{3/4}}\nonumber\\
[2mm]&& \times\, \left(1 +
\frac{k^2}{3(\sqrt{k^2+M^2}+M)^2}\right)\;.
\end{eqnarray}

%Don't forget to give each section, subsection,
%subsubsection, and paragraph a unique label (see
%Sect.~\ref{sec-1}).

\section{Parameters of the model}
\label{sec-3}

For the calculation of the  $\pi$- and $\rho$-meson
characteristics basing on the relations
(\ref{ffpi})--(\ref{decayconst111}) we use the following model
wave functions (see, e.g. \cite{ChC88pl, CoP05, Sch94}).

1. The Gaussian or harmonic oscillator wave function
\begin{equation}
u(k)= N_{HO}\, \hbox{exp}\left(-{k^2}/{2\,b_{\pi,\rho}^2}\right).
\label{HOwf}
\end{equation}

2. The power-law wave function:
\begin{equation}
u(k) =N_{PL}\,{(k^2/b_{\pi,\rho}^2 + 1)^{-n}}\;,\quad n=2,3\;.
\label{PLwf}
\end{equation}
In eqs. (\ref{HOwf}), (\ref{PLwf}) $b_{\pi,\rho}$ are parameters
of wave functions for pion and $\rho$-meson, respectively.

The electromagnetic form factors of constituent quarks in
(\ref{g_0}), (\ref{GqGRIP}) are taken in the form
\cite{KrT09,KrT98tmph}:
$$
G^{q}_{E}(Q^2) = e_q\,f_q(Q^2)\;,
$$
\begin{equation}
G^{q}_{M}(Q^2) = (e_q + \kappa_q)\,f_q(Q^2)\;, \label{q ff}
\end{equation}
where $e_q$ is the quark charge and $\kappa_q$ is the quark
anomalous magnetic moment,
\begin{equation}
f_q(Q^2) = \frac{1}{1 + \ln(1+ \langle r^2_q\rangle Q^2/6)}\; ,
\label{f_qour}
\end{equation}
$\langle r^2_q\rangle$ is the MSR of the constituent quark. Values
of all parameters used in these expressions are taken from the
$\pi$-meson calculation, see e.g.\ Ref.~\cite{TrT13, TrT15}.

So, the following parameters enter our calculations:

1) the parameters that describe the constituent quarks {\it per
se} (the quark mass $M$, the anomalous magnetic moments of the
quarks $\kappa_q$, that enter our formulae through the sum $s_q =
\kappa_u + \kappa_{\bar d}$, and the quark mean square radius
(MSR) $\langle r^2_q\rangle$);

2) the parameters  $b_{\pi,\rho}$ that enter the quark wave
functions (\ref{HOwf}), (\ref{PLwf}) and is determined by the
quark interaction potential.

In the paper \cite{KrT01} on pion we have shown that in our
approach all the parameters of the first group are the functions
of the quark mass $M$ and are defined by its value. In particular,
for the quark MSR we can use the relation (see, also
\cite{Car94}):
\begin{equation}
\langle r^2_q\rangle \simeq 0.3 /M^2\;. \label{r2q}
\end{equation}

To calculate electroweak properties of   the $\rho$ meson, we use
the same values of quark parameters from the first group as that
we have used for the pion \cite{KrT01}. So, the wave function
parameters $b_{\pi,\rho}$ are the only free parameters in our
calculations.

\section{Asymptotics of the pion form factor at high momentum transfers}
\label{sec-4}

It is worth to consider the pion form factor asymptotic behavior
(\ref{ffpi}) at $Q^2\to\infty$ especially. In our paper
\cite{KrT98tmph} it was shown that in our approach, the pion
form-factor asymptotics at
\begin{equation}
Q^2\to\infty\;,\quad M(Q^2)\;\to\; 0\;
 \label{limQM}
\end{equation}
does not depend on the choice of a wave function but is defined by
the relativistic kinematics only. We consider the fact that the
asymptotics obtained in our nonperturbative approach does coincide
with that predicted by QCD as a very significant one. Our approach
occurs to be consistent with the asymptotic freedom, and this
feature surely distinguishes it from other nonperturbative
approaches.

Let us note that it is obvious that at very high momentum
transfers the quark mass decreases as it goes to zero at the
infinity. Our approach permits to take into account the dependence
$M(Q^2)$ beginning from the range where this becomes necessary to
correspond to experimental data. It is possible that this will
take place at the values of $Q^2$ lower than $6$ GeV$^2$. So, it
is obtained in Ref. \cite{KrT98tmph} that
\begin{equation}
F_\pi(Q^2)\sim Q^{-2}\;. \label{Fpias}
\end{equation}

The asymptotics of the pion electromagnetic form factor $F_{\pi}$
at momenta transfer $Q^{2}\to \infty$ has been determined
\cite{FarrarJackson, EfremovRadyushkinAs, LepageBrodskyAs}, in the
QCD frameworks, as
\begin{equation}
Q^{2} F_{\pi}(Q^{2}) \to 8\pi \alpha_{\rm s}^{\rm 1-loop}(Q^{2})
f_{\pi}^{2}, \label{Eq:as}
\end{equation}
where $\alpha_{\rm s}^{\rm 1-loop}(Q^{2})=4\pi/\left(\beta_{0}
\log\left(Q^{2}/\Lambda_{\rm QCD}^{2} \right) \right)$ is the
one-loop running strong coupling constant, $\beta_{0}=11-2N_{f}/3$
is the first beta-function coefficient, $N_{f}$ is the number of
active quark flavours and $f_{\pi}\approx 130$~MeV  is the pion
decay constant. It is important to note that this asymptotical
behaviour, consistent with the quark counting rules
\cite{Qcounting1, Qcounting2}, includes the one-loop coupling only
and is to be modified whenever the one-loop approximation fails,
but not by means of a simple replacing of $\alpha_{\rm s}$ with
its more precise value. Involved QCD calculations have been
performed to obtain corrections to Eq.~(\ref{Eq:as}), see e.g.\
\cite{Bakulev?}. The QCD does not predict the value of $Q^{2}$ at
which this asymptotics should be reached.

\begin{figure}
\centering
\includegraphics[width=0.95\columnwidth]{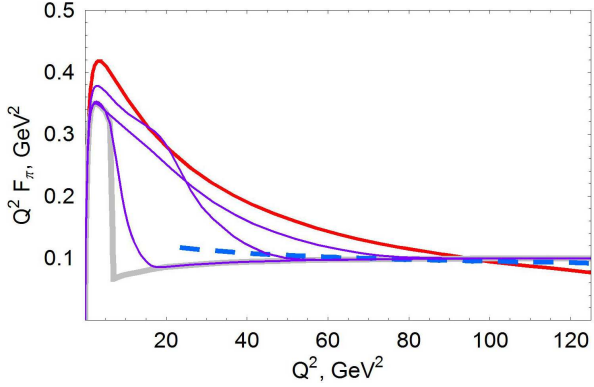}
\caption{Examples of the allowed solutions for $F_{\pi}(Q^{2})$
(thin lines) demonstrating how the QCD asymptotics,
Eq.(\ref{Eq:as}) (dashed line) settles down. The thick gray line
bounds from below the range of all solutions allowed by the
experimental constraints. The thick full (red) line represents the
solution with $M=$const, Refs.~\cite{KrT01, KrT09prc}.}
\label{fig:Fpi125GeV}
\end{figure}

The results of the calculation of $F_{\pi}$ from Ref. \cite{TrT13}
are presented in Fig.~\ref{fig:Fpi125GeV}.
As can be seen the our asymptotic coincides with QCD prediction at
different scenarios for the zero limit of constituent mass
(\ref{limQM}).

\section{The pion form factor in the region of the JLab experiments}
\label{sec-5}

\begin{figure}
\begin{center}
\includegraphics[width=0.95\columnwidth]{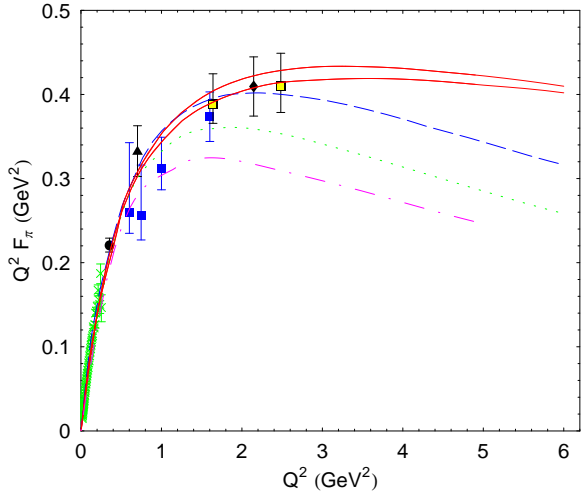}
\end{center}
\caption{ \label{fig2} Our predictions for the pion form factor
given in 1998~\cite{KrT01} (full red lines; upper line: wave
functions (\ref{HOwf}), lower line: wave functions (\ref{PLwf})
with $n=3$) compared with data and with some other models.
Experimental data are from Refs. \cite{Ack78, Bra79, Ame84, HuB08,
Blo08}. Other theoretical curves are: those by C.-W.~Hwang
\cite{Hwa01} (blue dashed); Cardarelli {\it et al.} \cite{Car94}
and (precisely coinciding with it) instant-form predictions
\cite{HeJ04} (magenta dash-dotted); Ref.~\cite{ChJ99} (green
dotted). Predictions of an upgraded version of a seminal paper
\cite{ChC88pl} (Ref.~\cite{CoP05}, Fig.~5, $M = 0.22$\, GeV) coincide
precisely with our upper curve.}
\end{figure}

The results of the calculation of the charge pion form factor
using the wave functions (\ref{HOwf}), (\ref{PLwf}) and the value
of constituent-quark mass $M$ = 0.22\,GeV (this parameter has been
fixed as early as in 1998 \cite{KrT01}  from the data at $Q^2 \le
0.26$ (GeV)$^2$ \cite{Ame84}) are shown in Fig. \ref{fig2}.

Let us note that our RQM describes well the experimental data for
the pion form factor including the recent points \cite{HuB08}. The
upper of our curves corresponds to the model (\ref{HOwf}), the
lower - to the models (\ref{PLwf}) with $n= 3$.

Let us emphasize that the parameters used in our calculations were
obtained from the fitting to the experimental data up to
$Q^2\;\simeq$ 0.26 GeV$^2$ \cite{Ame84}. At that time the data for
higher $Q^2$ was not correlated in different experiments and had
significant uncertainties. The later data for pion form factor in
JLab experiments up to $Q^2$ =2.45 GeV$^2$ were obtained with
rather good accuracy. All experimental points obtained in JLab up
to now agree very well with our prediction of 1998.

So, our way of fixing the model parameters constrains effectively
the behavior of wave functions both at small and at large relative
momenta. The structure of our relativistic integral representation
(\ref{ffpi}) is so, that  the form--factor behavior in the region
of small momentum transfers is determined by the wave function at
small relative momenta, and the behavior of the form factor in the
region of high momentum transfer --- by the wave function at large
relative momenta. The constraints for the wave functions provide
the limitations for the form factor, and this is seen in the
results of the calculation.

\section{Lepton decay constant and  MSR of the $\rho$-meson}
\label{sec-7}

Let us describe the procedure of calculation of the $\rho$- meson
MSR in detail, starting from the quark parameter, $M=0.22$~GeV,
used in a successful calculation of the pion parameters
\cite{KrT01}. As it has been demonstrated in Ref. \cite{KrT01}
(see also \cite{KrP15arxiv}), the actual choice of the
wave-function form does not affect the result provided the quark
parameters are fixed. In what follows, we illustrate the procedure
with the wave function (\ref{PLwf}) with $n=3$.

\begin{figure}
\begin{center}
\includegraphics[width=0.95\columnwidth]{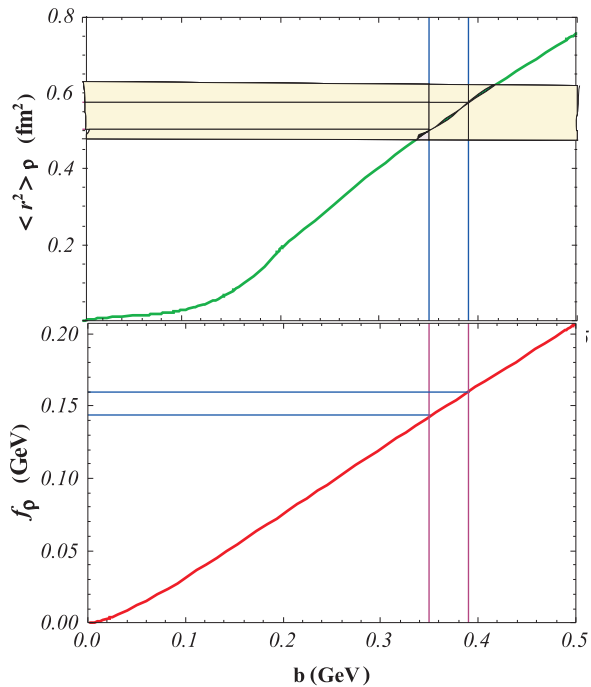}
\end{center}
\caption{ \label{constanta} The decay constant $f_{\rho}$ and the
$\rho$-meson MSR, $\langle r^2_\rho\rangle$,  as functions of the
only free parameter of the model, $b_\rho$. The experimental data
fix the value of $f_{\rho}$, as shown in the lower panel. This
fixes the value of $b_{\rho}$, which determines in turn the value
of  $\langle r^2_\rho\rangle$, as shown in the upper panel. The
values of quark parameters are taken from Ref.~\cite{KrT01}; the
wave function (\ref{PLwf}) with $n=3$ is used. }
\end{figure}

In the lower panel of Fig.~\ref{constanta},
the $\rho$-meson lepton decay constant $f_\rho$ as a function of
the only free parameter of the model, $b_\rho$, is presented. The
interval on the vertical axis representing the experimental values
of $f_\rho$, that is $f_{\rho}^{\rm exp} = (152 \pm 8)$~MeV
\cite{MeS15,Oli14}, is shown. It corresponds to the interval of
the values of $b_\rho$ which give, through our calculation, the
correct experimental values of the decay constant. This interval,
$b_\rho = (0.385 \pm 0.019)$~GeV, is shown on the horizontal axis
of Fig.~\ref{constanta}.

The calculated MSR of the $\rho$ meson is presented in the upper
panel of Fig.~\ref{constanta}. The interval of admissible values
of $b_{\rho}$ gives now the corresponding interval of MSR
predicted in the present study,
%that our approach predicts:
$ \langle r^{2}_\rho \rangle=(0.56 \pm 0.04)$~fm$^2$.

Table~\ref{tab:4} presents a comparison of our results with the
results of calculations of electroweak properties of the $\rho $
meson  in other approaches.
\begin{table}
\caption{The lepton decay constant $f_{\rho}$ and MSR of the
$\rho$ meson calculated within different approaches.}
\label{tab:4}
\begin{tabular}{ccc}
\hline
Model  & $f_{\rho}$, MeV& $\langle r^2_{\rho} \rangle$, fm$^2$ \\
\hline
{\bf This work}  & 152$\pm$8  & 0.56$\pm 0.04$  \\
& (fixed) & \\
\hline
\cite{MeS15} & 154 & 0.268 \\
\hline
\cite{BhM08} & 146 & 0.54 \\
\hline
\cite{RoB11} & 130  & 0.312 \\
\hline
\cite{CaB15} & --- & 0.67 \\
\hline
\cite{LoM00} & ---  & 0.49 \\
\hline
\cite{ChA15} & 147.4 & --- \\
\hline
\cite{OwK15} & --- & 0.67 \\
\hline
\cite{GrR07} & ---  & 0.655\\
\hline
\cite{MeS02} & ---  & 0.33 \\
\hline
\cite{CaG95} & ---  &  0.35\\
\hline
\cite{BaC02} & 134  & 0.296 \\
\hline
\cite{ChC07} & 133  & --- \\
\hline
\end{tabular}
\end{table}

The values of $\langle r^2_{\rho} \rangle$, while not measured
directly, are important for testing various conjectures about
strongly interacting systems. One of the interesting related
prediction was introduced as a consequence of the so-called
Wu--Yang hypothesis~\cite{WuY65} (see also Refs.~\cite{ChY68,
PoH87, PoH90, Gou74}), though it is remarkable by itself. Namely,
one may define the radius of a hadron either in terms of the
electroweak interaction (the mean square charge radius, $\langle
r^2_{\rm ch} \rangle$, calculated for the $\rho$ meson in this
paper) or in terms of the strong interaction (this radius,
$\langle r^2_{\rm st} \rangle$, is defined by the slope of the
cross section of hadron--proton scattering). The
conjecture~\cite{PoH90}, which may be derived from, though not
necessary implies, the hypothesis of Ref.~\cite{WuY65}, is the
equality of the two radii,
\begin{equation}
\langle r^2_{\rm st}\rangle = \langle r^2_{\rm ch}\rangle\;.
\label{rstch}
\end{equation}
This remarkable equality between two physical properties of a
hadron related to two different interactions of the Standard Model
has been verified experimentally with a great degree of accuracy
for the proton, $\pi$ and $K$ mesons.

\begin{figure}
\begin{center}
\includegraphics[width=0.95\columnwidth]{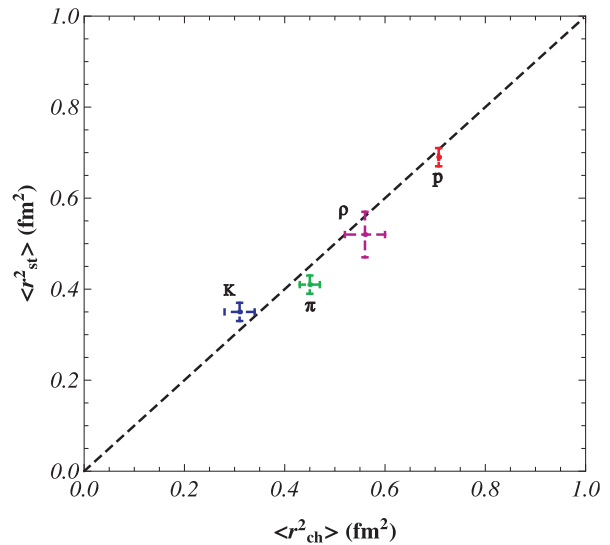}
\end{center}
\caption{ \label{radius} Relation between the strong-interaction
hadronic radius $\langle r^2_{\rm st}\rangle$ and the charge
radius $\langle r^2_{\rm ch}\rangle$ for light hadrons.}
\end{figure}

Even more demonstrative is Fig.~\ref{radius}, analogous to a
figure from the paper \cite{PoH90}, but presenting  more recent
data.
We can see that the value of the $\rho$-meson charge radius
obtained in this paper fits perfectly the conjecture
(\ref{rstch}).

\section{Analytic continuation of the pion form factor to the complex plain of momentum transfers}
\label{sec-8}

As mentioned above, one unsolved theoretical problem of RQM is its
relations to the fundamental QFT principles. It is currently
unknown whether the basic RQM axioms can be derived from the QFT
principles. This paper partially addresses this issue. We compare
the corollaries of the QFT principles with the model-independent
corollaries of the RQM axioms. In Ref. \cite{KrN13} we compared
the predictions for analytic properties of the pion form factor in
the complex plane of transferred momenta that follow from QFT
general principles with those obtained by the analytic
continuation of the form factor integral representation in the
spacelike domain derived in the framework of the instant form of
RQM \cite{KrT02}. We note that the problem of analytically
continuing the form factor in the RQM framework is first
formulated in the paper \cite{KrN13}.

Technically, the problem of constructing the pion form factor in
the complex plane of transferred momenta reduces to continuing
expression (\ref{ffpi}) analytically from the negative part of the
real axis to the complex plane of the parameter $t=-Q^2$.

We can show that the properties of the analytic continuation of
expression (\ref{ffpi}) depends strongly on the choice of the
constituent wave function. We must therefore find which
conditions, for example, to impose on the wave functions in order
to obtain the experimentally observed resonance behavior of the
form factor in the timelike domain of transferred momenta, i.e.,
on the positive part of the real axis of the parameter $t$.

It follows from the QFT microcausality condition that the pion
form factor is an analytic function in the complex plane of the
parameter t with a cut running from $4m_\pi^2$ to $\infty$, where
$m_\pi$ is the pion mass (see, e.g., \cite{BeD11, AnC12} and the
references therein). In this section, we show that our formulation
of the composite quark model has analogous properties. As noted
above, this fact is interesting from the standpoint of clarifying
the relation between the fundamental QFT approach and
phenomenological composite quark models.

The obtained analytic properties of the form factor differ from
those obtained in the QFT approach only by the position of the
branch point on the real axis: it is $4M^2$ in our case and
$4m_\pi^2$ in the QFT approach.

The main experimentally observed qualitative feature of the form
factor behavior in the complex plane is the presence of two
resonances, which correspond to $\rho$ and $\omega$ mesons and are
located close to each other on the positive part of the real axis
(in the timelike domain). Any solution of the problem of
constructing the form factor in the complex plane of transferred
momenta must ensure the existence of a resonance on the positive
part of the real axis for the analytically continued form factor.
We can demonstrate that wave functions in the momentum
representation without poles in the momentum complex plane do not
yield the resonance behavior of the pion form factor for timelike
transferred momenta. In particular, this is the case for a wave
function of the Gaussian type widely used in composite models
(see, e.g., \cite{ChC88pl,Sch94}), and this result holds
independently of the model parameter values. We therefore
encounter the problem of finding a wave function that produces
resonances and establishing the relation between the locus of the
resonance in the pion form factor and the locus of the wave
function poles. In solving this problem, the main idea is to
introduce wave functions with poles such that poles of form factor
(\ref{ffpi}) appear on the nonphysical sheet. Below, we show that
this ensures the desired resonance behavior of the form factor on
the positive part of the real axis.

So, the wave functions $u(k)$ (\ref{phi}) resulting in the
resonance behavior of the pion form factor for timelike
transferred momenta must have a pole at the point $k_s$:
$$
k_s^2 = \frac{M}{4}(z_s-2M)\;,\quad
$$
\begin{equation}
z_s = z_s'' + iz_s' = z_s'' + i\sqrt{t_r-4M^2}\;,\quad
z_s''\;<\;0\;, \label{u(ks)}
\end{equation}
where $t_r$ is resonance location.

Let the function $u(k)$ (\ref{phi}) be even and satisfies
condition of reality ($\hbox{Im}\,u(k)=0$) at real $k$ also.

The simplest rational function satisfying all the above conditions
is
\begin{equation}
u_{M}(k)=N\left[(k^2 - {k_s}^2)\,(k^2 -
({k_s}^*)^2)\right]^{-n}\;. \label{uM}
\end{equation}

The obtained analytic continuation with wave functions of form
(\ref{uM}) therefore contains the following parameters: $t_r$ and
$z''$ from (\ref{u(ks)}), the constituent mass $M$, and the
exponent $n$ of the wave function. We fix these data by fitting
experimental data for the parameters of the $\rho$ resonance in
the pion form factor, namely, its height, width, and location.

To describe the experimental data pertaining to the pion form
factor, we use the fitting by the celebrated Breit-Wigner formula
(see \cite{BrT08}):
\begin{equation}
F_{\pi BW}(t)=\frac{4\kappa^2}{4\kappa^2-t-2i\kappa\Gamma}\;,
\label{FBW}
\end{equation}
where $\kappa=0.375$ GeV, $\Gamma=0.1$ GeV. (we take the
parameters from \cite{BrT08}).

We numerically tuned the parameters of our model such that the
absolute value of the form factor best agrees with the
Breit-Wigner approximation of the experimental data, which gives
$$
t_r =\;0.57\,\mbox{CeV}^2\;,\quad z''_s = - 0.01\;,\quad
$$
\begin{equation}
M = 0.32\, \mbox{GeV}\;,\quad n = 0.86\;. \label{paramval}
\end{equation}

We note that the parameter $t_r$ is the resonance mass, and its
value is very close to the value $0.5625\, \mbox{GeV}^2$ provided
by formula (\ref{FBW}); the value of the constituent quark mass is
typical for different formulations of the composite quark model in
which constituent masses are in the interval $(0.2 -
0.33)\,\mbox{GeV}$ (see, e.g., \cite{KrT01, ChC88pl, Sch94,
Car94}). The parameter $z''_s$  determining the position of the
form factor singularity on the nonphysical sheet and the wave
function parameter $n$ are purely model parameters determined by
the tuning procedure.
\begin{figure}
\begin{center}
\includegraphics[width=0.95\columnwidth]{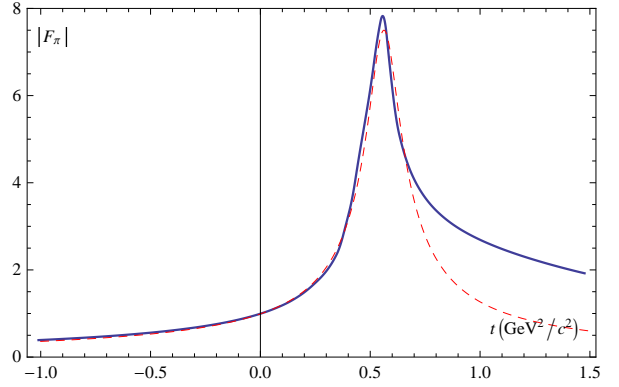}
\end{center}
\caption{The function $|F_{\pi}(t)|$ from: the solid curve is
expression (\ref{ffpi}), and the dashed curve is the Breit-Wigner
approximation.} \label{1}
\end{figure}

The results of numerically integrating expression (\ref{ffpi})
with wave function (\ref{uM}) and of calculating with the
Breit-Wigner formula are shown in Fig. \ref{1}. We can see that
our analytic continuation of pion form factor (\ref{ffpi}) from
the spacelike domain to the complex plane provides a good
qualitative description of the form factor behavior in the
timelike domain already for the simplest wave function choice
given by (\ref{uM}). The deviation from the Breit-Wigner formula
(or from the experimental data) in the domain of the squared
transferred momentum above the maximum point is presumably because
our choice (\ref{uM}) of the wave function is too robust. We note
that the choice of the wave function only weakly affects the form
factor description in the spacelike domain; for a more precise
description of the pion form factor in the timelike domain of
transferred momenta, we need a more refined expression for the
wave function, for instance, with a larger number of poles.

\begin{figure}
\begin{center}
\includegraphics[width=0.95\columnwidth]{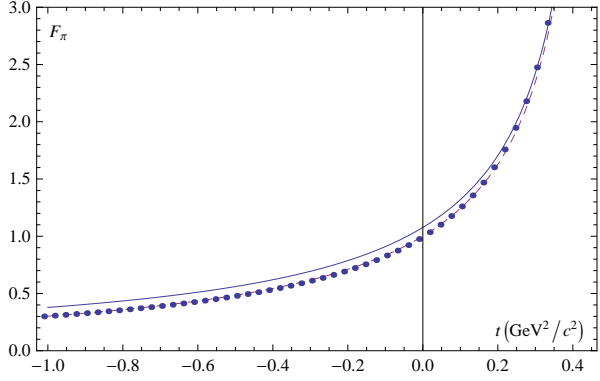}
\end{center}
\caption{Comparing the calculated pion form factor with the known
dispersion relations: points correspond to a straightforward
calculation with formula (\ref{ffpi}), the solid line is obtained
from dispersion relation (\ref{dire1}), and the dashed line is
obtained from dispersion relation (\ref{dire2}). The results of
calculating by formula (\ref{ffpi}) practically coincide with
those obtained using dispersion relation (\ref{dire2}).} \label{2}
\end{figure}

We can demonstrate the correctness of our constructed analytic
continuation as follows. We first verify how well the constructed
form factor satisfies the known dispersion relation in the
spacelike domain
\begin{equation}
F_{\pi}(t) = \frac{1}{\pi}\int\limits_{4M^2}^{\infty}
\frac{Im\left[F_{\pi}(t')\right]}{t'-t}dt'\;, \label{dire1}
\end{equation}

and also the dispersion relation with one subtraction (with the
weakened condition of decreasing at infinity)
\begin{equation}
F_{\pi}(t)=1 + \frac{t}{\pi}\int\limits_{4M^2}^{\infty}
\frac{Im\left[F_{\pi}(t')\right]dt'}{t'(t'-t)}\;. \label{dire2}
\end{equation}
In Fig. \ref{2}, we compare our result with formulas (\ref{dire1})
and (\ref{dire2}). We use parameter values (\ref{paramval}). It
can be seen that the constructed pion form factor in the complex
plane satisfies the dispersion relations with good accuracy.

We also verified the correctness of the proposed analytic
continuation against the condition that the $\pi\pi$-scattering
$S$-matrix is unitary, which can be written in terms of the
elastic pion form factor in the form (see, e.g., \cite{ShS67, KaK14}):
\begin{equation}
\hbox{Im}\,F_\pi(t) =
F_\pi(t)\,\exp\left(-i\delta^1_1(t)\right)\,\sin\,\delta^1_1(t)
\;, \label{Unitary}
\end{equation}
where $\delta^1_1(t)$ is $\pi\pi$-- scattering phase at the state
$I = J = 1$.

\begin{figure}
\begin{center}
\includegraphics[width=0.95\columnwidth]{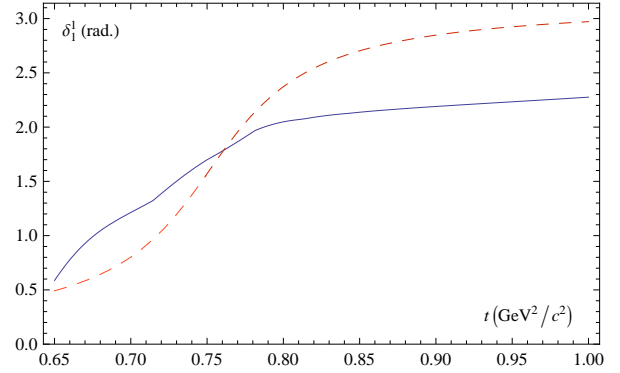}
\end{center}
\caption{The $\pi\pi$-scattering phase: the dashed line is the
phase calculated by the Breit-Wigner formula, and the solid line
is the result of phase calculation using formula (\ref{ffpi}).}
\label{3}
\end{figure}

We present our calculation results for the $\pi\pi$-scattering
phase in Fig. \ref{3}. It can be seen that our constructed form
factor provides a scattering phase description close to the
results of calculations based on the Breit-Wigner formula.

As already noted, the fitting based on the Breit-Wigner formula
satisfactorily describes experimental data for the pion form
factor. The problem of a more detailed description of the phase
and character of the resonance behavior of the pion form factor in
the framework of our approach is related to the problem of
choosing a wave function of interacting quarks that is more
realistic than (\ref{uM}). Calculation results depend strongly on
this choice, which ensures the possibility of a more precise
description of the experimental data.

\section{Conclusions}
\label{sec-9}

The approach developed in IF RQM has following main features:
predictivity: parameters of model are fixed from experimental data
at small momentum transfers. The experimental data for the pion
form factor at large momentum transfers obtained later are
described without tuning of parameters; robustness: the behavior
of the pion and $\rho$-meson electromagnetic form factors does not
depend on the choice of wave functions and are determined by the
mass of the constituent quarks; the approach gives the asymptotics
that agrees with QCD asymptotic behavior at large momentum
transfers; the approach gives the self-consistent description of
the electroweak properties of the pion and $\rho$-meson; the
approach gives the right analytical properties of the pion form
factor in the complex plane of momentum transfers.

\begin{acknowledgement}
One of the authors (VT) thanks Sergey Troitsky for interesting
discussions. Authors (AK and RP) thank the Organizing committee
for his kind invitation to the XIX International Seminar on High
Energy Physics "QUARKS-2016" and hospitality. This work was
supported in part (AK and RP) by the Ministry of Education and
Science of the Russian Federation (grant No. 1394, state task).
\end{acknowledgement}

\end{document}